\documentclass[aps,preprint,preprintnumbers,nofootinbib,showpacs]{revtex4}
\usepackage{graphicx,color}
\begin{document}
\title{Split Supersymmetry, stable gluino, and gluinonium}
\author{Kingman Cheung}
\affiliation{
Department of Physics and NCTS, National Tsing Hua University, Hsinchu, 
Taiwan, R.O.C.
}
\author{Wai-Yee Keung}
\affiliation{
Department of Physics, University of Illinois at Chicago, 
Chicago IL 60607-7059, U.S.A.
}
\date{\today}

\begin{abstract}
In the scenario recently proposed by Arkani and Dimopoulos, the supersymmetric
scalar particles are all very heavy, at least of the order of $10^9$ GeV but
the gauginos, higgsinos, and one of the neutral Higgs bosons remain under a
TeV.  The most distinct signature is the metastable gluino.
However, the detection of metastable gluino depends crucially on 
the spectrum of hadrons that it fragments into.  Instead, here we 
propose another unambiguous signature by forming the gluino-gluino bound 
state, gluinonium, 
which will then annihilate into $t\bar t$ and $b\bar b$ pairs. 
We study the sensitivity of such signatures at the LHC.
\end{abstract}
\preprint{}
\maketitle

\section{Introduction}
Supersymmetry (SUSY) is one of the most elegant solutions, if not the best, 
to the gauge hierarchy problem. It also provides a dynamical mechanism 
for electroweak symmetry breaking, as well as a viable candidate for 
dark matter (DM).  However, the fine-tuning arguments restrict  
SUSY to be in TeV scale, otherwise the fine tuning problem
returns. Theorists for the past two decades have been plugging holes
that the weak scale SUSY model may fall into.  
Recently, Arkani and Dimopoulos \cite{arkani} 
adopted a rather radical approach to 
SUSY, essentially they discarded the hierarchy problem and accepted the
fine-tuning solution to the Higgs boson mass. They argued that the 
much more serious problem, the cosmological constant problem, needs a much
more serious fine-tuning that one has to live with, and so why not let go
of the much less serious one, the gauge hierarchy problem.  

Once we accept this proposal the finely-tuned Higgs scalar boson is not a 
problem anymore.  The more concern issue is to a find a consistent set of
parameters so as to satisfy the observations.  (i) The result of the
Wilkenson Microwave Anisotropy Probe (WMAP) \cite{wmap} has refined 
the DM density to be $\Omega_{\rm DM} h^2 = 0.094 - 0.129$ ($2\sigma$ range),
(ii) neutrino mass, and (iii) cosmological constant.  The last one is accepted
by the extremely fine-tuned principle.   The second observation requires
heavy right-handed neutrinos of mass scale of $10^{11-13}$ GeV, so that it
does not have appreciable effects on electroweak scale physics.  The first
observation, on the other hand, requires a weakly interacting particle of
mass $\alt 1$ TeV, in general.  It is this requirement which affects most of
the parameter space of the finely-tuned SUSY scenario.  

Such a finely-tuned SUSY scenario was also named 
``split supersymmetry'' \cite{giudice},
which can be summarized by the followings.
\begin{enumerate}
\item All the scalars, except for a CP-even Higgs boson, are all super heavy
that their common mass scale $\tilde{m} \sim 10^{9}$ GeV 
$ - M_{\rm GUT}$.
The scenario 
will then be safe with flavor-change neutral currents, CP-violating 
processes, e.g., EDM.

\item The gaugino and the higgsino masses are relatively much lighter
and of the order of TeV, because they are protected by a $R$ symmetry and 
a PQ symmetry, respectively.

\item A light Higgs boson, very similar to the SM Higgs boson.

\item The $\mu$ parameter is relatively small due to the requirement of DM.
This is to make sure that there are sufficient mixings in the neutralino
sector such that the lightest neutralino can annihilate efficiently to give
the correct DM density.

\item This scenario still allows gauge-coupling unification.

\end{enumerate}

An important difference between split SUSY and the usual MSSM is, as already
pointed out by a number of authors \cite{arkani,giudice,aaron}, that the 
gaugino-Higgsino-Higgs couplings are no longer the same as the corresponding
gauge couplings at energies below the SUSY breaking scale $M_{\rm SUSY}$,
although they are the same at the 
scale $M_{\rm SUSY}$ and above.  We need to evolve the gaugino-Higgsino-Higgs 
couplings
down from $M_{\rm SUSY}$ 
using the renormalization group equations involving only
gauginos and Higgsinos.

The most distinct feature for this scenario in terms of collider 
phenomenology is that the gluino now becomes
metastable inside collider detectors, because all sfermions are super heavy.
This feature is very similar to the gluino-LSP scenario \cite{gunion,raby}
as far as gluino collider signatures are concerned.
The gluino pair so
produced will hadronize into color-neutral hadrons by combining
with some light quarks or gluons.
Such objects are strongly-interacting massive particles,
electrically either neutral or charged.  If the hadron is
electrically neutral, it will pass through the tracker with little
trace.  
If the hadron is electrically charged, it will undergo ionization
energy loss in the central tracking system, hence behaves like a
``heavy muon''. 
However, an issue arises when the neutral hadron containing the
gluino may convert into a charged hadron when the
internal light quark or gluon is knocked off and replaced by another
light quark. And vice versa.
The probability of such a scattering depends crucially
on the mass spectrum of the hadrons formed by the gluino and light quarks 
and gluons.  In reality, we know very little about the spectrum.
Some previous estimates \cite{gunion} assumed a fixed probability that
the gluino fragments into a charged hadron.  The resulting sensitivity 
depends crucially on this probability.

In view of such an uncertainty,  
we propose to look at another novel
signature of stable or metastable gluinos.  Since the gluinos are produced 
in pairs and stable, they can form a bound state, 
called gluinonium\cite{keungkhare,haber}
by exchanging gluons.
In fact, one could talk about the toponium were not if the decay time of the
top quark is too short.  The potential between two massive gluinos can be
very well described by a coulombic potential.  The value of the
wave function at the origin can be reliably determined.  
We will calculate the production rates for the gluinonium at the LHC and
at the Tevatron.  

The two gluinos within the gluinonium can then annihilate into standard model
particles, either $gg$ or $q\bar q$, depending on the angular momentum
and color of the bound state.  Since each gluino is in a color octet 
$\textbf{8}$, the gluinonium can be in 
\begin{equation}
\textbf{8} \otimes \textbf{8} = \textbf{1} + \textbf{8}_S + \textbf{8}_A 
      + \textbf{10} + \overline{\textbf{10}} + \textbf{27} \;,
\end{equation}
where the subscript ``$S$'' stands for symmetric and ``$A$'' 
for antisymmetric. 
It was shown that only $\textbf{1}$, $\textbf{8}_S$, and $\textbf{8}_A$ 
have an attractive potential \cite{haber}.  
Here we only consider $S$-wave bound states with a total spin $S=0$ or $1$.
In order that the total wave function of the gluinonium is antisymmetric,
the possible configurations are $^1S_0(\textbf{1})$, $^1S_0(\textbf{8}_S)$,
and $^3S_1(\textbf{8}_A)$, the color representations of which are shown in the
parentheses.  Note that the dominant decay mode of 
$^1S_0(\textbf{1})$ and $^1S_0(\textbf{8}_S)$ is $gg$, which would give rise
to dijet in the final state.  However, the QCD dijet background could easily
bury this signal and we have verified that.
Therefore, we focus on the $^3S_1(\textbf{8}_A)$ state,
which decays into $q\bar q$ including light quark and heavy quark pairs.
We could then make use of the $t\bar t$ and/or $b \bar b$ in the final state
to search for the peak of the gluinonium in the invariant mass spectrum of the 
$t\bar t$ and/or $b\bar b$ pairs.  
Nevertheless, we shall give the production cross sections for both
$^1S_0$ and $^3S_1$ states.

The study in this work is not just confined to split SUSY scenario, but also 
applies to other stable gluino scenarios, e.g., gluino-LSP \cite{gunion,raby},
long-lived gluino scenario, {\it etc}.
Our calculation shows that gluinonium production and its decay can help
searching for the stable or metastable gluinos.
Although the sensitivity using this signature is in general 
not as good as the ``heavy muon''-like signature, it is, however, 
free from the uncertainty of the hadronic spectrum of the gluino hadrons.

\section{Production of gluinonium}

In calculating gluinonium production we will encounter the spinor combination
$u(P/2) \bar v(P/2)$, where $P$ is the 4-momentum of the gluinonium.
We can replace it by, in the nonrelativistic approximation,
\begin{eqnarray}
^3S_1(\textbf{8}_A) &:& 
     u(P/2) \bar v(P/2) \longrightarrow \frac{1}{\sqrt{2}}
\frac{R_8(0)}{2\sqrt{4\pi M}}
\, \frac{1}{\sqrt{3}} f^{hab}\, \not\!\epsilon(P) \,(\not\!{P} + M )
 \nonumber   \\
^1S_0(\textbf{8}_S) &:& 
     u(P/2) \bar v(P/2) \longrightarrow \frac{1}{\sqrt{2}}
\frac{R_8(0)}{2\sqrt{4\pi M}}
\, \sqrt{\frac{3}{5}} d^{hab}\, \gamma^5 \,(\not\!{P} + M )  \label{11}\\
^1S_0(\textbf{1}) &:& 
     u(P/2) \bar v(P/2) \longrightarrow \frac{1}{\sqrt{2}}
\frac{R_1(0)}{2\sqrt{4\pi M}}
\, \frac{1}{\sqrt{8}} \delta^{ab}\, \gamma^5 \,(\not\!{P} + M ) \nonumber \;,
\end{eqnarray}
where the color factors $\frac{1}{\sqrt{3}}f^{hab}$, 
$\sqrt{\frac{3}{5}} d^{hab}$,
and $\frac{1}{\sqrt{8}} \delta^{ab}$ are the color representations of
$\textbf{8}_A$,  $\textbf{8}_S$, and $\textbf{1}$,  respectively.
The $\epsilon(P)$ is the polarization 4-vector for the gluinonium
of momentum $P$ .  
The values of the color octet and singlet wave functions at the origin 
are given by the coulombic potential between the gluinos, 
with one-gluon approximation \cite{haber},
\begin{eqnarray}
\left | R_8(0)\right |^2 &=& \frac{27 \alpha^3_s(M) M^3}{128} \;, \\
\left | R_1(0)\right |^2 &=& \frac{27 \alpha^3_s(M) M^3}{16} \;.
\end{eqnarray}
There is an additional factor of $1/\sqrt{2}$ in Eqs. (\ref{11}) because of 
the identical gluinos in the wave function of the gluinonium.  

\begin{figure*}[t]
\centering
\includegraphics[width=3in]{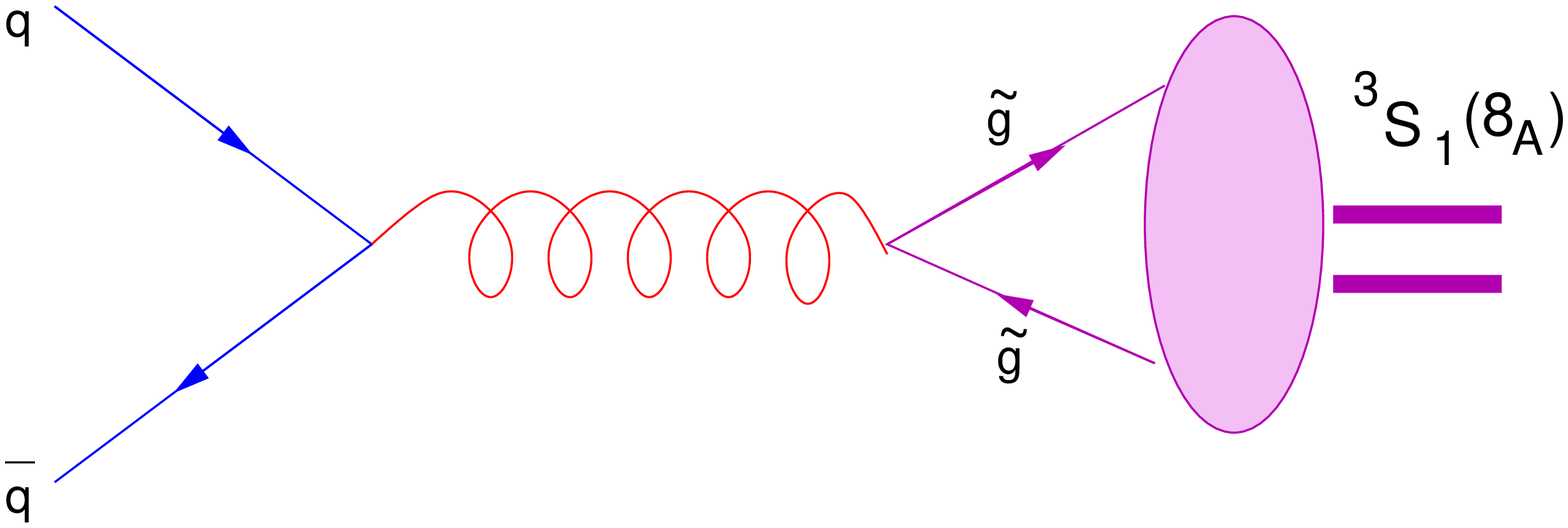}

\includegraphics[width=2in]{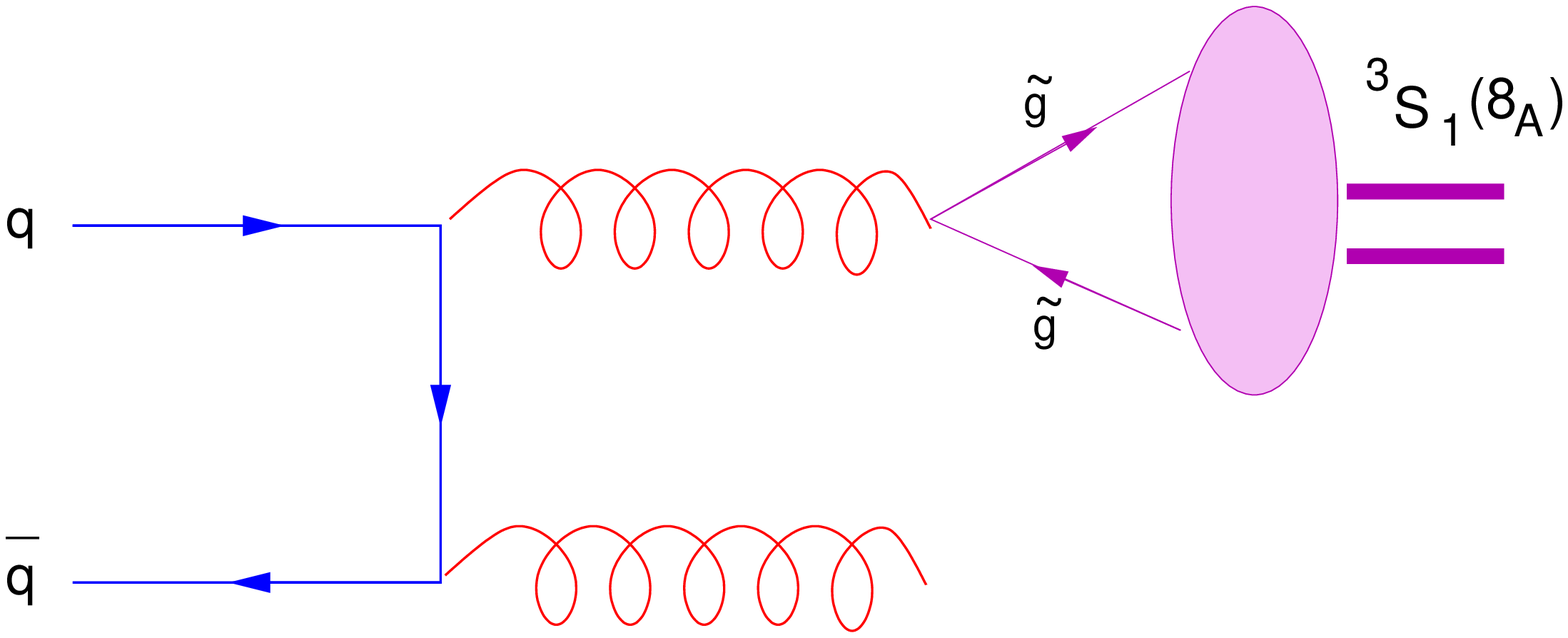}
\includegraphics[width=2in]{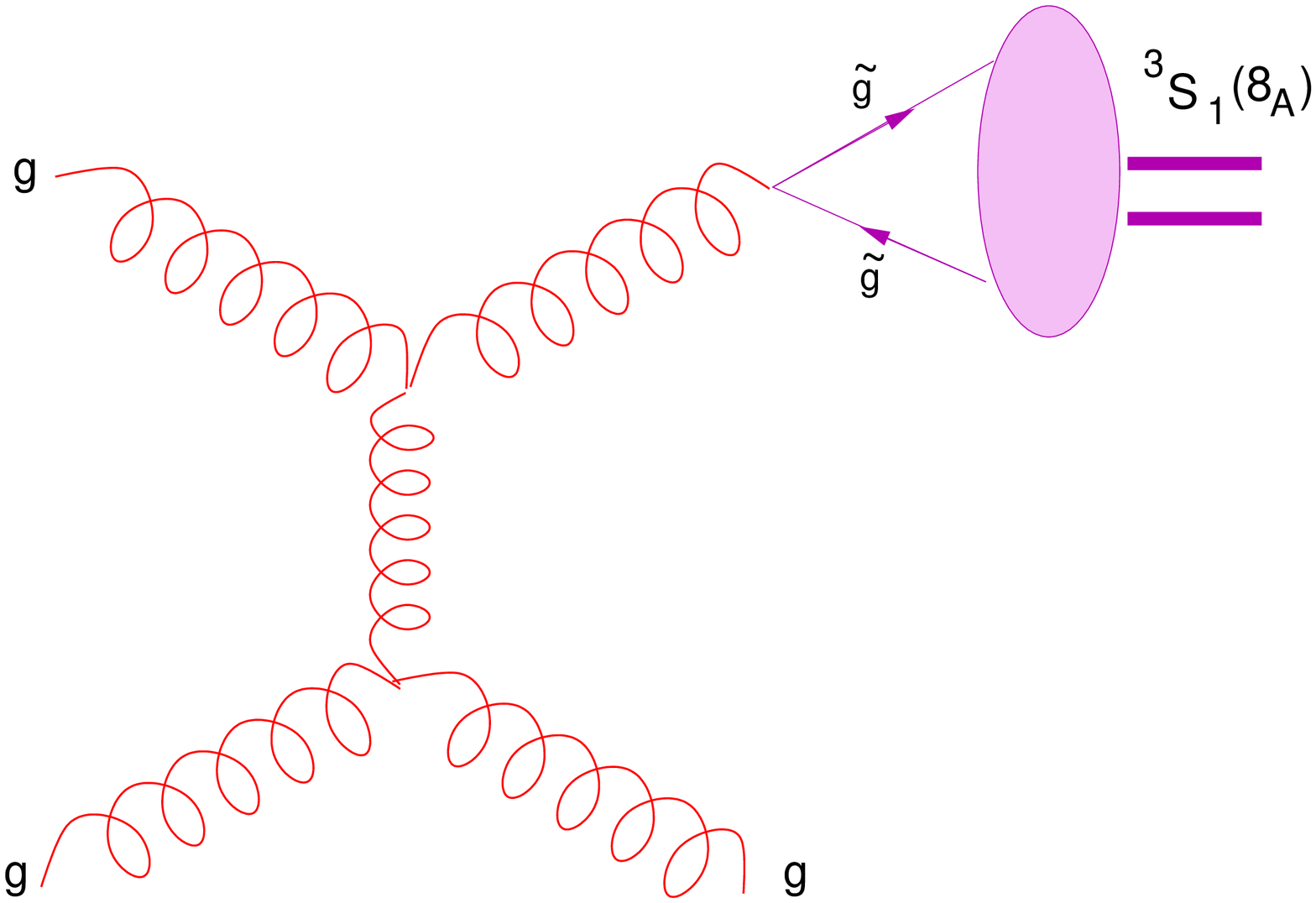}
\includegraphics[width=2in]{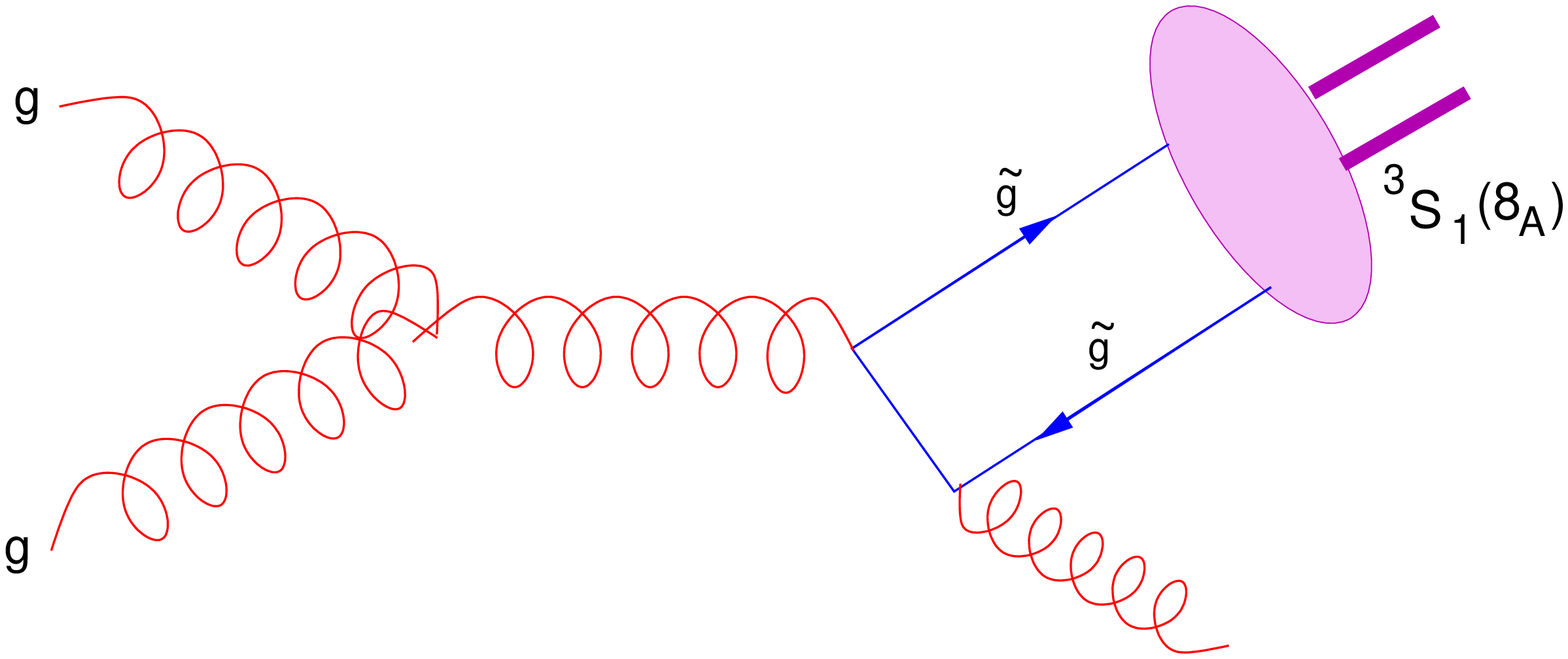}
\caption{\small \label{fey}
Representative Feynman diagrams for $q\bar q \to ^3\!S_1(\textbf{8}_A)$, 
$q \bar q \to ^3\!S_1(\textbf{8}_A ) +g$, and 
$gg  \to ^3\!S_1(\textbf{8}_A ) +g$.
}
\end{figure*}

In the calculation of $^3S_1(\textbf{8}_A)$, 
the lowest order process is a $2\to 1$ process:
\begin{equation}
q \bar q \to   ^3S_1(\textbf{8}_A ) \label{2-1} \;.
\end{equation}
The next order $2\to 2$ processes include 
\begin{eqnarray}
q \bar q & \to &   ^3S_1(\textbf{8}_A ) +g\label{2-2}  \\
q g & \to &   ^3S_1(\textbf{8}_A ) +q \label{2-3} \\
g g & \to &   ^3S_1(\textbf{8}_A ) +g \label{2-4}  \;
\end{eqnarray}
which would give a $p_T$ to the gluinonium.  
Representative Feynman diagrams are
shown in Fig. \ref{fey}.
Naively, one would expect the $gg$ fusion in Eq.~(\ref{2-4}) has a large
cross section because of the high $gg$ luminosity and fragmentation type
diagrams.  However, when other non-fragmentation type diagrams are included,
we found that the cross section is extremely small for heavy gluinos.  
On the other hand, the processes in Eqs. (\ref{2-2}) and (\ref{2-3}) 
give a small correction to the process in
Eq.(\ref{2-1}).  Neverthesless, one also has to consider similar or even 
larger corrections to the $t\bar t$ background.  Therefore, we only use
the lowest order process of Eq. (\ref{2-1}) and $t\bar t$ background
in the signal-background analysis.

The cross section for $q \bar q \to ^3S_1(\textbf{8}_A)$ is given by
\begin{equation}
\hat \sigma = \frac{16 \pi^2 \alpha_s^2}{3}\, \frac{|R_8(0)|^2}{M^4} \,
\delta(\sqrt{\hat s} - M) \;.
\end{equation}
After folding with the parton distribution functions, the total cross section
is given by
\begin{equation}
\sigma = \frac{32 \pi^2 \alpha_s^2}{3 s} \frac{|R_8(0)|^2}{M^3} \,
  \int f_{q/p}(x)      f_{\bar q/p} (M^2/s x) 
\frac{dx}{x} \;,
\end{equation}
where $s$ is the square of the 
center-of-mass energy of the colliding protons and we
sum over all possible initial partons $q=u,\bar u,d,\bar d, s,\bar s, c,
\bar c, b,\bar b$.

\begin{figure}[t!]
\centering
\includegraphics[width=5in]{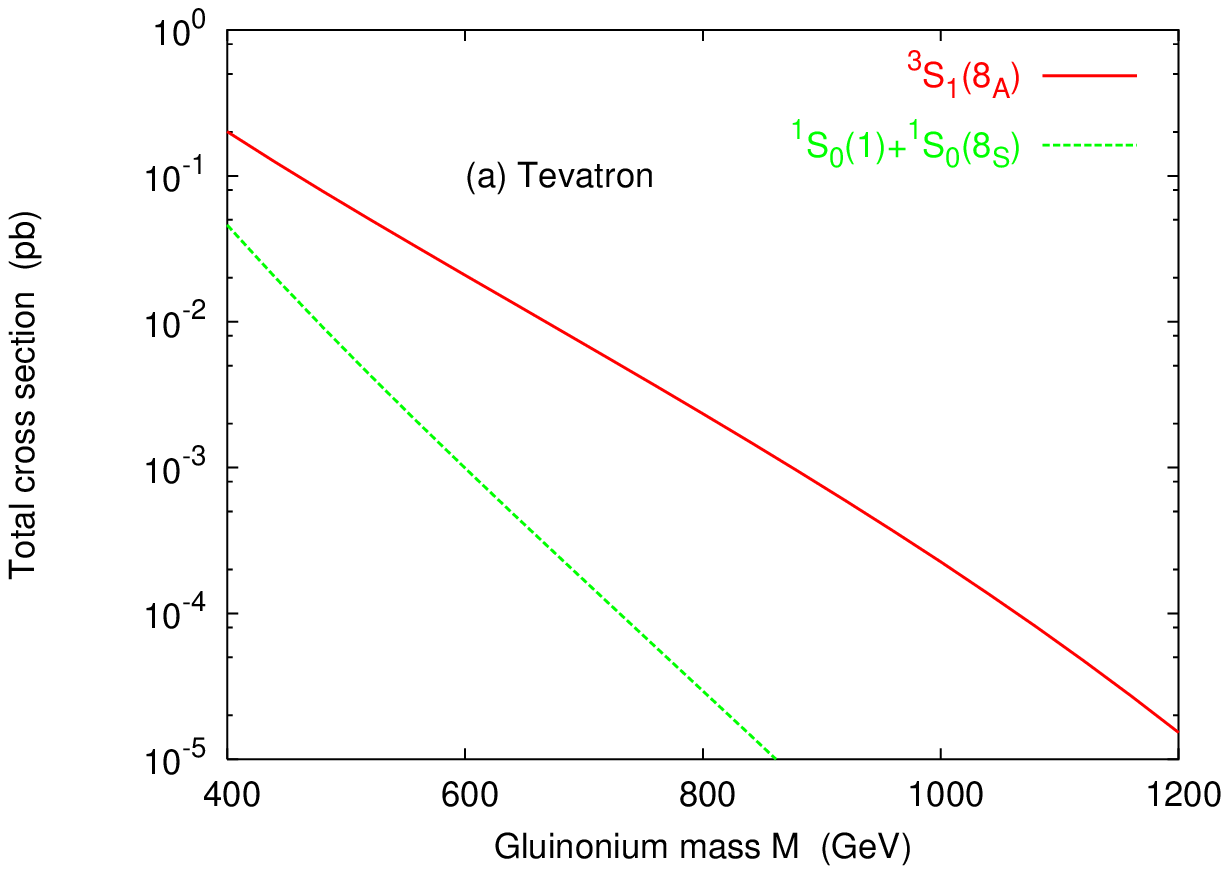}
\includegraphics[width=5in]{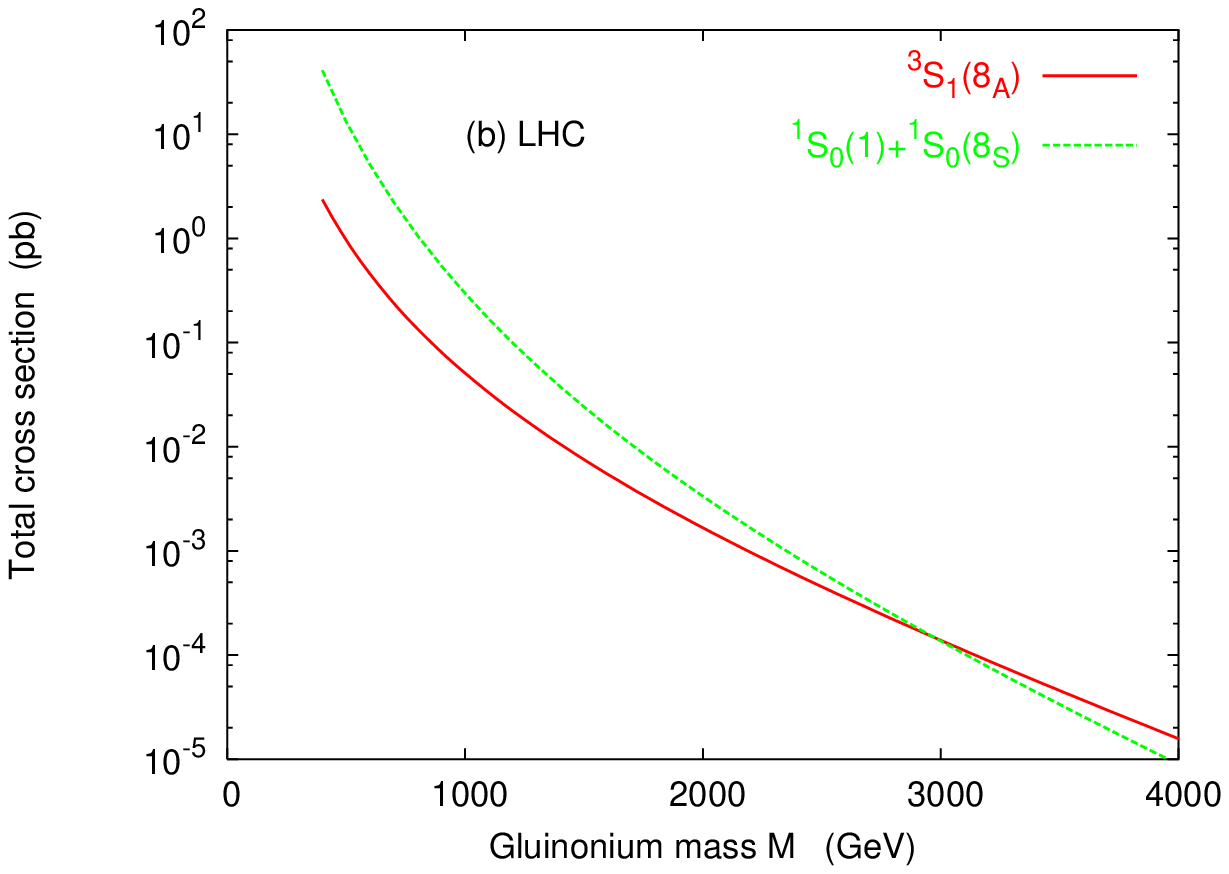}
\caption{\small \label{total}
Total production cross sections for gluinonium in $^3S_1(\textbf{8}_A)$ 
and in $^1S_0(\textbf{1}) + ^1S_0(\textbf{8}_S)$ states at the (a) Tevatron
and (b) LHC}
\end{figure} 

One can also estimate the decay width of the gluinonium by adding all partial
widths into $u\bar u,\; d\bar d, \; s\bar s,\; c \bar c,\; b\bar b,\; t\bar t$:
\begin{equation}
\Gamma\left(^3S_1(\textbf{8}_A) \right )
= \sum_{Q=u,d,s,c,b,t} \, \alpha_s^2 \frac{|R_8(0)|^2}{M^4}\,
 (M^2 + 2 m_Q^2) \, \sqrt{1 - 4m_Q^2/M^2 } \;.
\end{equation} 
When the gluinonium mass is 1 TeV or above, the branching ratio into 
$t\bar t$ is $1/6$.  
We note that the formulas for $^3S_1(\textbf{8}_A)$  production and decay
width differ from those given in Refs.~\cite{russians} that our results
are smaller by a factor of $2$.
 
For completeness we also give the formulas for $^1S_0(\textbf{1})$ and
$^1S_0(\textbf{8}_S)$:
\begin{eqnarray}
\Gamma(^1S_0(\textbf{1})) &=&18 \alpha_s^2 \frac{ |R_1(0)|^2}{M^2} \nonumber \\
\Gamma(^1S_0(\textbf{8}_S))&=& \frac{9}{2} \alpha_s^2 
   \frac{ |R_8(0)|^2}{M^2} \nonumber \\
\sigma(pp\to ^1S_0(\textbf{1})) &=& 
  \frac{9 \pi^2 \alpha_s^2}{4 s} \frac{|R_1(0)|^2}{M^3} \,
  \int f_{g/p}(x) f_{g/p} (M^2/s x) \frac{dx}{x}  \nonumber \\
\sigma(pp\to ^1S_0(\textbf{8}_S)) &=& 
  \frac{9 \pi^2 \alpha_s^2}{2 s} \frac{|R_8(0)|^2}{M^3} \,
  \int f_{g/p}(x) f_{g/p} (M^2/s x) \frac{dx}{x}  \;.
\end{eqnarray}
Note that our singlet $^1S_0(\textbf{1})$ production and decay width
formulas agree with Ref. \cite{haber}, but the 
octet $^1S_0(\textbf{8}_S)$ production and decay width
formulas differ from those in Ref. \cite{haber} that our results
are smaller by a factor of $8$.

We show the production cross section for $^3S_1(\textbf{8}_A)$ and
$^1S_0(\textbf{1})$ and $^1S_0(\textbf{8}_S)$ at the Tevatron and the LHC
in Fig. \ref{total}.  Note that we have included a factor of $\zeta(3)
\simeq 1.2$ because of the sum of all radial excitations 
$\sum_n (1/n^3)$ \cite{haber}.  The strong coupling constant is evaluated
at the scale of gluinonium mass $M$ at the one-loop level.
At the Tevatron, the $q\bar q$ luminosity dominates and so the vector
gluinonium is more important.  On the other hand, 
at the LHC, the pseudoscalar gluinonium has larger
production cross sections until $M\approx 3$ TeV, 
because of the large $gg$ initial parton luminosity at small $x$.
However, it is noted that the 
dijet background is far more serious than the $t\bar t$ background, and
so in the next section we focus on $^3S_1(\textbf{8}_A)$ signal and
$t\bar t$ background.

\section{Background Analysis}
The $^3S_1(\textbf{8}_A)$ gluinonium decays into light quark and heavy quark
pairs.  The signal of light quark pairs would be easily buried by QCD dijet 
background.  Thus, we focus on top-quark pair.  Irreducible backgrounds comes
from QCD $t\bar t$ production.  
We take the advantage that the 
$t$ and $\bar t$ coming from the heavy gluinonium would have a very large 
$p_T$, typically of order of the mass of the gluino, while the 
$t$ and $\bar t$ from the background 
are not.  We impose a very large $p_T$ cut as follows
\begin{eqnarray}
p_T(t), \; p_T(\bar t) & > & \frac{3}{4} \, m_{\tilde{g}} \;\;\; \mbox{for 
$M \ge 1$ TeV} \nonumber \\
p_T(t), \; p_T(\bar t) & > & 100 \;{\rm GeV} \label{cut} \;\;\; \mbox{for
$M < 1$ TeV} \;.
\end{eqnarray}
The background has a continuous spectrum while the signal plus background
should show a small bump right at the gluinonium mass, provided
that the signal is large enough.
Since the intrinsic width of the
gluinonium is very small, of the order of 1 GeV, the width of the observed 
bump is determined by experimental resolution.  We have adopted a simple
smearing of the top quark momenta: $\delta E/E = 50\%/\sqrt{E}$.  
We summarize the cross sections at the LHC 
for the signal and background
in Table \ref{table1}.  One could also include using $b\bar b$ in the final
state.  The branching ratios of the gluinonium into $b\bar b$ and 
$t\bar t$ are the same for such heavy gluinonia.  Naively, one would
expect the QCD background of $b\bar b$ production to be roughly the same
as $t\bar t$ production at such high invariant mass region.  Therefore, 
by including $b\bar b$ in the final state, although one does not improve 
the signal-background ratio, one would, however, improve the 
significance $S/\sqrt{B}$ of the signal by a factor of $\sqrt{2}$.
{}From the Table the sensitivity is only up to about $M=2 m_{\tilde{g}}=0.5-
0.6$ TeV with a luminosity of 100 fb$^{-1}$.


\begin{table}[th!]
\caption{\small \label{table1}
Cross sections at the LHC for the gluinonium signal into $t\bar t$
with mass $M$ and 
the continuum $t\bar t$ background 
between $M-50$ GeV  and $M+50$ GeV.  If including $b\bar b$ in the final
state the significance $S/\sqrt{B}$ would increase by a factor 
of $\sqrt{2}$.}
\medskip
\begin{ruledtabular}
\begin{tabular}{cccc}
$M = 2m_{\tilde{g}}$   &  $\sigma(^3S_1(\textbf{8}_A))$
                            & $t\bar t$ bkgd (fb)     
                            & $S/\sqrt{B}$ \\
(TeV)  & (fb) & (fb) &    for $L=100$ fb$^{-1}$ \\
\hline
$0.5$  & $120$   & $83000$   & $4.2$ \\
$0.75$ & $28$    & $19000$   & $2.0$ \\
$1$    & $4.9$   & $1150$ & $1.4$ \\
$1.5$  & $0.78 $ & $97  $ & $0.79$  \\
$2.0$  & $0.17 $ & $14  $ & $0.45$  \\
$3.0$  & $0.014$ & $0.67$ & $0.17$ 
\end{tabular}
\end{ruledtabular}
\end{table}

The $t\bar t$ background has some uncertainties due to higher order
corrections, structure functions, reconstruction of top quark momenta, etc.
Similar corrections also apply to the signal.
The uncertainty of the signal calculation lies
in the determination of $|R_8(0)|^2$, which should be small due to the good
approximation of coulombic potential between heavy gluinos.

\section{Comparison with ``heavy muon'' signature }

Another important signature of stable or metastable gluinos 
is that once gluinos are produced they will hadronize into 
color-neutral hadrons by combining with some light quarks or gluons.
Such objects are strongly-interacting massive particles,
electrically either neutral or charged.  If the hadron is
electrically neutral, it will pass through the tracker with little
trace.  However, 
an issue arises when the neutral hadron containing the
gluino may convert into a charged hadron when the
internal light quark or gluon is knocked off and replaced by another
light quark. The probability of such a scattering depends crucially
on the mass spectrum of the hadrons formed by the gluino and light quarks 
and gluons.
In reality, we know very little about the spectrum, so we simply
assume a 50\% chance that a gluino will hadronize into
a neutral or charged hadron.
If the hadron is electrically charged, it will undergo ionization
energy loss in the central tracking system, hence behaves like a
``heavy muon''. 
Essentially, the penetrating particle
loses energy by exciting the electrons of the material.
Ionization energy loss $dE/dx$
is a function of $\beta \gamma \equiv p/M$ and the charge $Q$ of the
penetrating particle \cite{pdg}.  
For the range of
$\beta\gamma$ between 0.1 and 1 that we are interested in, $dE/dx$
has almost no explicit dependence on the mass $M$ of the
penetrating particle. Therefore, when $dE/dx$ is measured in an
experiment, the $\beta\gamma$ can be deduced, which then gives the
mass of the particle if the momentum $p$ is also measured.  Hence,
$dE/dx$ is a good tool for particle identification for massive
stable charged particles.
In addition, one can demand the massive
charged particle to travel to the outer muon chamber and deposit energy 
in it.  To do so the particle must have at least a certain initial 
momentum depending on the mass; typical initial $\beta\gamma = 0.25 - 0.5$.
In fact, the CDF Collaboration has made a few searches for massive
stable charged particles \cite{cdf1}. The CDF analysis required
that the particle produces a track in the central tracking chamber
and/or the silicon vertex detector, and at the same time
penetrates to the outer muon chamber.
The CDF requirement on $\beta\gamma$ is 
\[
0.26-0.50 \;\; \alt \;\;\; \beta \gamma \;\;\; < \;\; 0.86  \;.
\]
We use a similar analysis for metastable gluinos.
We employ the following acceptance cuts on the gluinos
\begin{equation}
\label{cuts}
p_T(\tilde{g}) > 20 \;{\rm GeV}\,,\;\; |y(\tilde{g})|<2.0\,,
\;\; 0.25 < \beta\gamma < 0.85\;.
\end{equation}
It is easy to understand that a large portion of cross section satisfies
the cuts; especially the heavier the gluino the closer to the threshold 
is.  In 
Table~\ref{stable} we show the cross sections from direct
gluino-pair production with all the
acceptance cuts in Eq. (\ref{cuts}), for detecting 1 massive
stable charged particle (MCP), 2 MCPs, or at least 1 MCPs in the
final state.   The latter cross section is the simple sum of the
former two.  We have used a probability of $P=0.5$
that the $\tilde{g}$ will hadronize into a charged hadron.
Requiring about 10 such events as suggestive
evidence, the sensitivity can reach up to about
$m_{\tilde{g}}\simeq 2.25$ TeV with an integrated luminosity 
of 100 fb$^{-1}$.
In the Table \ref{stable} we also show the cross sections of 
$\sigma_{\geq \rm 1MCP}$ for $P=0.1$ and $P=0.01$ in the last two columns.
As expected the cross section decreases with $P$, the probability that
the gluino hadronizes into a charged hadron.
If the probability is only of order of 10\%, the sensitivity will be
slightly less than 2 TeV. If the probability is of order of 1\%,
the sensitivity will go down to 1.5 TeV.
This is the major uncertainty associated with gluino detection using the
method of stable charged tracks.
Note that the treatment here is rather simple.  For more 
sophisticated detector 
simulations please refer to Refs. \cite{gunion,kilian,rizzo}.

\begin{table}[t!]
\caption{\small Cross sections for direct gluino-pair production at the 
LHC, with the cuts of Eq.~(\ref{cuts}).
Here $\sigma_{\rm 1MCP}$, $\sigma_{\rm 2MCP}$, and 
$\sigma_{\geq \rm 1MCP}$ denote requiring the
detection of 1, 2, and at least 1 massive stable charged particles (MCP) in the
final state, respectively.  Here the probability $P$ that gluinos fragment
into charged states is $P=0.5$.  We also show the cross section 
$\sigma_{\geq \rm 1MCP}$ for $P=0.1$ and $P=0.01$ in the last two columns.
\label{stable} }
\medskip
\begin{ruledtabular}
\begin{tabular}{llllll}
 $m_{\tilde{g}}$ (TeV) & $\sigma_{\rm 1MCP}$ (fb)
  & $\sigma_{\rm 2MCP}$ (fb) & $\sigma_{\geq \rm 1MCP}$ (fb) 
& $\sigma_{\geq \rm 1MCP}$  (fb) 
& $\sigma_{\geq \rm 1MCP}$  (fb) \\
 & $P=0.5$  & $P=0.5$ & $P=0.5$ & $P=0.1$ & $P=0.01$ \\
\hline
$0.5$ & $4050$  & $620$ & $4670$   & $1040$ & $105$ \\
$1.0$ & $67$    & $13$ & $80$      & $18$   & $1.9$ \\
$1.5$ & $3.7$   & $0.91$ & $4.6$   & $1.1$  & $0.11$ \\
$2.0$ & $0.3$   & $0.09$ & $0.39$  & $0.09$ & $0.0096$ \\
$2.25$& $0.092$ & $0.029$ & $0.12$ & $0.029$ &$0.003$   \\
$2.5$ & $0.028$  & $0.0095$ & $0.038$ & $0.009$ & $0.0009$ 
\end{tabular}
\end{ruledtabular}
\end{table}

\section{Conclusions}
The most distinct signature for split SUSY or gluino-LSP scenario
is that the gluino is stable or metastable within the detector.  Previous 
studies are based on hadronization of the gluino into $R_{\tilde{g}}$ 
hadrons, but however the detection of such a signature has a large
uncertainty due to the unknown spectrum of $R_{\tilde{g}}$ hadrons.
We have demonstrated that using the gluinonium is free from
this uncertainty, and the decay of gluinonium into a $t\bar t$ and/or
$b\bar b$ pair may
provide a signal above the continuum $t\bar t$ invariant mass spectrum.
However, a rather good resolution of $t\bar t$ spectrum and accurate
determination of continuum background are necessary to bring out the
signal.

{\it Note added}: during writing other papers on split
SUSY appear \cite{zhu,anch,indians}.

\section*{Acknowledgments}
This research was supported in part by
the National Science Council of Taiwan R.O.C. under grant no.
NSC 92-2112-M-007-053- and 93-2112-M-007-025-, and in part by the US
Department of Energy under grant no. DE-FG02-84ER40173.
W.-Y. K. also thanks the hospitality of NCTS in Taiwan while this work was
initiated.

\end{document}